\documentclass[aps,pra,twocolumn,superscriptaddress]{revtex4-1}
\pdfoutput=1
\usepackage{amssymb,amsmath,amsfonts,amsbsy,amsthm,soul,bm}

\usepackage{dsfont}
\usepackage{bbold}

\usepackage{graphicx}
\usepackage{hyperref}
\hypersetup{
 colorlinks=true,
 linkcolor=blue,
  filecolor=magenta,      
  urlcolor=cyan,
}

\def\<{\langle}
\def\>{\rangle}
\def\dag{\dagger}

\newcommand{\tr}{\text{Tr}}
\newcommand{\ket}[1]{| #1 \rangle}
\newcommand{\bra}[1]{\langle #1 |}

\newcommand{\var}{\text{Var}}
\newcommand{\cov}{\text{Cov}}

\begin{document}
\title{An approximate description of quantum states}

\author{Marco Paini}
\email{mpaini@rigetti.com}
\affiliation{Rigetti Computing, 138 Holborn, London, EC1N 2SW, UK.\vspace{0.2em}}

\author{Amir Kalev}
\email{amirk@umd.edu}
\affiliation{Joint Center for Quantum Information and Computer Science, University of Maryland, College Park, MD 20742-2420, USA.\vspace{1em}}

\date{\small\today}

\begin{abstract}
We introduce an approximate description of an $N$--qubit state, which contains sufficient information to estimate the expectation value of any observable with precision independent of $N$. We show, in fact, that the error in the estimation of the observables' expectation values decreases as the inverse of the square root of the number of the system's identical preparations and increases, at most, linearly in a suitably defined, $N$--independent, seminorm of the observables. Building the approximate description of the $N$--qubit state only requires repetitions of single--qubit rotations followed by single--qubit measurements and can be considered for implementation on today's Noisy Intermediate-Scale Quantum (NISQ) computers. The access to the expectation values of all  observables for a given state leads to an efficient variational method for the determination of the minimum eigenvalue of an observable. The method represents one example of the practical significance of the approximate description of the $N$--qubit state. We conclude by briefly discussing extensions to generative modelling and with fermionic operators.
\end{abstract}

%%%%%%%%%%%%%%%%%%%%%%%%%%%%%%%%%%%%%%%%%%%%%%
%                                                      INTRODUCTION                                                               %
%%%%%%%%%%%%%%%%%%%%%%%%%%%%%%%%%%%%%%%%%%%%%%

\maketitle

\section{Introduction}
Quantum density operators represent our knowledge of the state of quantum systems and gives us a way to calculate expectation values and predict experimental results via the Born rule. Our knowledge of a system's density operator is equivalent to our ability to calculate the expectation value of any observable of the system; unfortunately, even for a finite $N$--body system, reconstructing the system's density operator requires the knowledge of exponentially many numbers in $N$. This exponentially--expensive representation translates into practical difficulties in estimating and storing density operators, in the form of density matrices, even of systems composed of just a few tens of qubits.

Over the last few years, there has been a surge of interest in devising protocols to reduce the resources required to estimate and store density operators~\cite{Aaronson2007learnability,Vidal2003Efficient,Gross2010Quantum,Kalev2015Quantum,Aaronson2018Shadow,Gosset2019compressed}. Compressed sensing tools were used~\cite{Gross2010Quantum,Kalev2015Quantum} in the context of quantum state tomography, the canonical quantum state reconstruction protocol, to reduce the number of measurements necessary for robust estimation of density matrices. The theory of Probably Approximately Correct (PAC) learning was utilised in~\cite{Aaronson2007learnability} to show that if, instead of full tomography, we are interested in approximating with high probability the expectation values of observables drawn from a distribution, then only polynomially--many samples in $N$ of the observables' expectation values are required. It was later also shown~\cite{Aaronson2018Shadow} that estimating the expectation values of only a few given observables can be achieved with approximation using only polynomially--many copies in $N$ of the system. As for storing, it is well established that quantum states that can be represented as low bond--dimension matrix product states can be efficiently stored on classical computers~\cite{Vidal2003Efficient}. A more recent result showed that quantum states can be approximately described using information obtained from an order of $\sqrt{2^N}$ inner products with stabiliser states~\cite{Gosset2019compressed}. The growth of the number of copies of the system in $N$ for estimation and the considerable requirements in $N$ for storing the information associated to a generic quantum state lead to the question of whether a protocol that significantly reduces or eliminates the dependence on $N$ is admissible. In this work, we give an affirmative answer to the question.

Instead of focusing on the estimation and storage of the density matrix, we start by considering the equivalent problem of calculating the expectation value of any observable of the system up to a certain precision. Infinite--dimensional quantum systems, such as one mode of the electromagnetic field, have density matrices with infinitely--many entries. Nonetheless, we know that we can calculate expectation values of several observables with arbitrary precision only requiring a finite number of copies of the system~\cite{dmp, dlp,tokyo,raymer95,sch}. Therefore, unless finite--dimensional spaces have a property that is not reflected in infinite dimension, for a finite $N$--qubit system, we would expect not only a non--exponential growth in $N$ of the number of copies required to achieve a given precision in the calculation of observables' expectation values, but, at most, a bounded growth of the number copies in $N$, if a dependence of the number of copies in $N$ is present at all. On the other hand, the methods of homodyne tomography~\cite{dmp, dlp,tokyo} reveal a dependence of the required number of copies, or equivalently of the statistical errors, on the properties of the observables of interest, in particular, on the observables' boundedness. Following the lead of these observations, we show in this article that, for a system of $N$ qubits, the expectation value of any observable can be calculated with arbitrary precision with a number of copies of the system independent of $N$, but dependent on a seminorm of the observable. The information obtained through the measurements on the copies of the system will be used to form an approximate description of the quantum state -- a description that does not suffer from the exponential growth in $N$ of density matrices and that, in principle, can be constructed and stored even for systems larger than today's NISQ computers. The exact definition of the observables' seminorm and of the approximate description of the quantum state will be given in Sections \ref{SU(2) Tomography} and \ref{approximate description}. The mathematical framework that will be utilised is based on group theory tomography introduced in~\cite{Paini2000Quantum}, which will be briefly explained in the rest of this section.

\par
\vskip 1em

The system we will consider specifically is a system of $N$ qubits; however, the considerations that will be made are in general applicable to systems with different Hilbert spaces, infinite--dimensional, too. Even if ultimately we are interested in the expectation values of the observables, we will start by manipulating the density operator. Irreducible representations of groups provide explicit forms for a basis in the space ${\cal B}(\cal H)$ of linear bounded operators defined in the finite--dimensional space $\cal H$ of our system. Any operator $A\in{\cal B}(\cal H)$ can be expressed as
\begin{align}
A = \int_{\cal G} d\mu(g)\,{\mathrm {Tr}}\big[A\,{T}(g)\big]{T}^{\dag}(g),              \label{general tomo}
\end{align}
where $\cal G$ is a compact Lie group, $d\mu(g)$ is Haar’s invariant measure for $\cal G$ multiplied by ${\rm {dim}}(\cal H)$ and $T$ an irreducible, unitary ray representation of $\cal G$. Choosing the density operator $\rho$ as $A$ and calculating the trace in (\ref{general tomo}) over the eigenvectors of the operators $T(g)$ leads to
\begin{align}
\rho=\int_\Lambda d\mu(\lambda)\sum_{t}p(t,\lambda)K(t,\lambda),							\label{tomo essence}
\end{align}
where $\int_\Lambda d\mu(\lambda)$ is over all the classes ${\cal G}_\lambda$ of $\cal G$ defined by containing elements corresponding to operators $T(g)$ with the same eigenvectors, $\sum_{t}$ is over the eigenvalues $t=t(g)$ of $T(g)$, $K(t,\lambda)$ is known as kernel operator and $p(t,\lambda)$ represents the probability that a measurement of $T(g_\lambda)$ gives $t$, with $g_\lambda$ equal to any element $g$ of ${\cal G}_\lambda$ for a given $\lambda$.

If we take the trace of both sides of (\ref{tomo essence}) after multiplication by a generic observable $O$, Eq. (\ref{tomo essence}) becomes
\begin{align}
\langle O \rangle=\int_\Lambda d\mu(\lambda)\sum_{t}p(t,\lambda){R}[O](t,\lambda),		\label{expectation values}
\end{align}
with the estimator ${R}[O](t,\lambda)$ defined as ${\mathrm {Tr}}[OK(t,\lambda)]$. The form of (\ref{expectation values}) suggests that an approximate value of $\langle O \rangle$ could be calculated with Monte Carlo methods~\cite{DAriano2003Quantum}. The random variable $O_M$ defined as
\begin{align}
O_M=\frac 1M\sum_{j=1}^M{R}[O](t_j, \lambda _j),							\label{monte carlo}
\end{align}
with $\lambda _j$ sampled uniformly in $\Lambda$ and $t_j$ measured value of $T(g_{\lambda_j})$, is asymptotically in $M\to\infty$ a normal variable with expectation value equal to $\langle O \rangle$ and variance
\begin{align}
\var(O_M)=\frac 1M\,\var({R}[O](t, \lambda)).							\label{variance}
\end{align}
Equation (\ref{monte carlo}) tells us that if we prepare our system $M$ times and each time we take a sample $\lambda_j$ from $\Lambda$, measure $T(g_{\lambda_j})$ obtaining $t_j$ and calculate ${R}[O](t_j, \lambda _j)$, we can approximate the expectation value of $O$ with Eq. (\ref{monte carlo}), with statistical error decreasing as $1/\sqrt{M}$. Clearly, the method is only useful if we can control $\var({R}[O](t, \lambda))$. The evaluation of the variance of the estimator will be discussed in the following section.

%%%%%%%%%%%%%%%%%%%%%%%%%%%%%%%%%%%%%%%%%%%%%%
%                                                      SU(2) Tomography                                                           %
%%%%%%%%%%%%%%%%%%%%%%%%%%%%%%%%%%%%%%%%%%%%%%

\section{SU(2) Tomography} \label{SU(2) Tomography}

%											One Qubit														%
\subsection{One qubit} \label{tomo 1 qubit}
The compact group $SU(2)$ admits an irreducible, unitary representation in every finite--dimensional space and we can choose the one in a space of dimension 2 for a 1--qubit system. The idea of $SU(2)$ tomography is certainly not new~\cite{Paini2000Quantum,DAriano2003Spin}. We will derive here the 1--qubit estimators and an upper bound for their variance, laying the foundations of the $N$--qubit system generalisation of the next subsection.

\par
\vskip 1em

With the parametrisation of $SU(2)$ with the angles $(\vartheta,\varphi,\psi)$ belonging to $[0,\pi]{\times}[0,2\pi){\times}[0,2\pi]$ and $\vec n$ defined as $(\cos \varphi\sin\vartheta,\sin\varphi\sin\vartheta,\cos\vartheta)$, the operators $T(\vec n,\psi)=e^{-i\psi\vec s\cdot\vec n}$, with the spin operator $\vec s$ equal to $1/2$ of Pauli operators, constitute an irreducible, unitary representation of $SU(2)$~\cite{cornwell}. Formula (\ref{tomo essence}) becomes~\cite{Paini2000Quantum,DAriano2003Spin}:
\begin{align}
\rho=\int_{\Sigma}\frac{d\vec n}{4\pi}\sum_{m_s=-1/2}^{1/2}p(m_s,\vec n)K_1(m_s,\vec n),				\label{tomo 1-qubit}
\end{align}
with the unitary spherical surface ${\Sigma}$ as integration domain, $p(m_s,\vec n)$ representing the probability that $m_s$ is the result of a measurement of $\vec s\cdot\vec n$ and the 1--qubit kernel operator $K_1(m_s,\vec n)$ given by
\begin{align}
K_1(m_s,\vec n)=\frac{2}{\pi}\int_0^{2\pi}d\psi\sin^2\frac\psi2 e^{i\psi(m_s-\vec s\cdot\vec n)}. 	\label{kernel}
\end{align} 
The 1--qubit estimator ${R}_1$ depends on $K_1(m_s,\vec n)$, but also on the choice of the observable $O$. Given the estimators ${R}[O]$ are linear in $O$, we can expand the generic observable $O$ on the identity $\mathbb{1}=\sigma_0$ and Pauli operators $\sigma_x=\sigma_1,\,\sigma_y=\sigma_2,\,\sigma_z=\sigma_3$ as
\begin{align}
O=\sum_{i=0}^{3}a_i\,\sigma_i 															\label{O}
\end{align}
and only have to calculate ${R}_1$ for the identity and the Pauli operators, as ${R}_1[O]$ will be given by
\begin{align}
{R}_1[O]=\sum_{i=0}^{3}a_i\,{R}_1[\sigma_i].											\label{R linear}
\end{align}
We show in Appendix \ref{app 1 qubit} that the estimators for the identity operator $\mathbb{1}$ and the Pauli operator $\sigma_\alpha$, with $\alpha=x,y,z$, are
\begin{align}
{R}_1[\mathbb{1}](m,\vec n)&=1, 						\label{identity estimator}\\ 
{R}_1[\sigma_\alpha](m,\vec n)&=3\,m\,n_\alpha,			\label{pauli estimator}
\end{align}
with $m=2\,m_s=\pm 1$ eigenvalues of the Pauli operators, and have variances
\begin{align}
\var({R}_1[\mathbb{1}](m,\vec n))&=0,					\label{identity variance}\\
\var({R}_1[\sigma_\alpha](m,\vec n))&\leqslant3,			\label{pauli variance}
\end{align}
where the last inequality results from the variance of a random variable being smaller than or equal to the expectation value of the square of the variable and $\langle {R}_1^2[\sigma_\alpha](m,\vec n) \rangle=3$. For the generic observable $O$, we introduce a seminorm $\Vert O\Vert$ in ${\cal B}({\cal H})$ defined by
\begin{align}
\Vert O\Vert^2=3\sum_{i=1}^{3}a_i^2, 										\label{norm 1 qubit}
\end{align}
where $\{a_i\}\in\mathds{R}$ are the coefficient of (\ref{O}). We prove in Appendix \ref{app 1 qubit} that
\begin{align}
\var({R}_1[O](m,\vec n))\leqslant{\Vert O\Vert}^2.						\label{O variance}
\end{align}

\par
\vskip 1em

To calculate $\langle O \rangle$ using (\ref{monte carlo}), we need to sample $(\vartheta,\varphi)$ uniformly from ${\Sigma}$ and measure $\vec s\cdot\vec n$. Since the eigenvectors $|\vec n,m\rangle$ of $\vec s\cdot\vec n$ can be obtained from the eigenvectors $|m\rangle$ of $s_z$ with $|\vec n,m\rangle=e^{-i\vartheta\vec s\cdot\vec n_\perp}|m\rangle$, where $\vec n_\perp=(-\sin\varphi,\cos\varphi,0)$, the measurement of $\vec s\cdot\vec n$ can be conveniently replaced with a measurement in the computational basis of the eigenvectors of $s_z$ with a rotation $e^{i\vartheta\vec s\cdot\vec n_\perp}$ before measurement. The tomographic procedure for 1--qubit becomes: sample $(\vartheta_1,\varphi_1)$, apply the rotation $e^{i\vartheta_1\vec s\cdot\vec n_{1_{\perp}}}$, measure in the computational basis, obtaining $m_1$, calculate ${R}_1[O](m_1,{\vec n}_1)$ using (\ref{R linear}), (\ref{identity estimator}) and (\ref{pauli estimator}). Repeat $M$ times obtaining ${R}_1[O](m_2,{\vec n}_2),\ldots,{R}_1[O](m_M,{\vec n}_M)$ and compute $O_M$ of (\ref{monte carlo}). The value thus calculated is equal to $\langle O \rangle$ with normal statistical error, for $M$ sufficiently large, smaller than or equal to ${\Vert O\Vert}/{\sqrt{M}}$.

%												N Qubits													%
\subsection{\boldmath{$N$} qubits}
The method developed for one qubit in the previous subsection is easily generalised to $N$ qubits, in the assumptions that the qubits are distinguishable and that a qubit-specific measurement capability is available~\cite{Paini2000Quantum,DAriano2003Spin}. The space of the $N$--qubit system is the tensor product of the single--qubit spaces and we can write (\ref{tomo essence}) for $N$ qubits choosing ${\cal G}=SU(2)^{N}$ and $T(g)=\bigotimes_{k=1}^N e^{-i{\psi}_k{\vec s}_k\cdot{\vec n}_k}$, where $g\in {\cal G}$ and $\times$ and $\bigotimes$ indicate the direct product of groups and the tensor product of operators respectively. Formula (\ref{tomo 1-qubit}) becomes
\begin{align}
\rho=&\bigg(\frac{1}{(4\pi)^N}\prod_{k=1}^N \int_{{\Sigma}_k}{d{\vec n}_k}\sum_{m_{s_k}}\bigg)p(m_{s_1},{\vec n}_1,\ldots,m_{s_N},{\vec n}_N)\nonumber\\
&\times K(m_{s_1},{\vec n}_1,\ldots,m_{s_N},{\vec n}_N),				\label{tomo N-qubit}
\end{align}
with $K(m_{s_1},{\vec n}_1,\ldots,m_{s_N},{\vec n}_N)$ equal to the tensor product of single--qubit kernels
\begin{align}
K(m_{s_1},{\vec n}_1,\ldots,m_{s_N},{\vec n}_N)=\bigotimes_{k=1}^N K_1(m_{s_k},{\vec n}_k)	\label{kernel N-qubit}
\end{align}
and $p(m_{s_1},{\vec n}_1,\ldots,m_{s_N},{\vec n}_N)$ representing the probability that a measurement of
${\vec s}_1\cdot{\vec n}_1,\ldots,{\vec s}_N\cdot{\vec n}_N$ gives $(m_{s_1},\ldots,m_{s_N})$. As a result of (\ref{kernel N-qubit}), the estimator ${R}$ for an observable given by the tensor product of single--qubit observables $O_k$ is simply the product of the single--qubit estimators:
\begin{align}
{R} \bigg[\bigotimes_{k=1}^N O_k\bigg]=\prod_{k=1}^N{R}_1[O_k].			\label{estimator product of single qubits estimators}
\end{align}
Before we consider the general form of an observable for the $N$--qubit system, let us consider the simple case of an observable given by a single--qubit observable $O_l$. If $N=1$, in other terms, if the single qubit represents the entire system, Eq. (\ref{R linear}) and (\ref{O variance}) give us the expectation value of $O_l$ with the associated statistical error. If the qubit was instead part of a larger system of qubits, we would calculate the expectation value of the observable $\mathbb{1}\otimes\cdots\otimes\mathbb{1}\otimes O_l \otimes\mathbb{1}\otimes\cdots\otimes\mathbb{1}$ using (\ref{estimator product of single qubits estimators}). Because of (\ref{identity estimator}) and (\ref{identity variance}), the estimator (\ref{estimator product of single qubits estimators}) would reduce to the single--qubit estimator. Put differently, the estimator and, as we will see, the tomographic procedure produce the same approximation for the expectation values of the same physical observables, for all dimensions of the Hilbert space of the system. To formalise this property and generalise the tomographic procedure for the $N$--qubit system, we start by writing the generic observable $O$ for the $N$--qubit system as
\begin{align}
O=\sum_i \,a_i\,\sigma_{i_1}\cdots\sigma_{i_N},											\label{general O}
\end{align}
with $i_k=0,\ldots,3$ for every $k$, $i=(i_1,\ldots\,i_N)$ and the tensor product sign in the string of Pauli operators having been omitted for a simpler notation. Using Eq. (\ref{estimator product of single qubits estimators}), the estimator for $O$ is given by
\begin{align}
{R}[O]=\sum_ia_i\prod_{k=1}^N{R}_1[\sigma_{i_k}].						\label{general estimator}
\end{align}
We show in Appendix \ref{app N qubits} that, as in (\ref{O variance}), the variance of the estimator has the upper bound
\begin{align}
\var({R}[O])\leqslant{\Vert O\Vert}^2,						\label{general variance}
\end{align}
with the seminorm defined as
\begin{align}
\Vert O\Vert^2=\sum_{i,j (\neq 0_v)}3^{r_{ij}}\Delta_{ij}|a_i||a_j|
 										\label{norm N qubits}
\end{align}
where the sum is extended to all values of $i$ and $j$ except $0_v\equiv(0,0,\ldots,0)$, $r_{ij}$ is the number of indices $k$ for which $i_k\neq 0 \wedge  j_k\neq 0$, $\Delta_{ij}=0$ if there exists a $k$ such that $i_k\neq 0 \wedge  j_k\neq 0 \wedge i_k\neq j_k$ and $\Delta_{ij}=1$ otherwise. One could argue that the presence in the seminorm of $3^{r_{ij}}$ reflects the exponential growth of the estimators' variance in $N$, as, for example, for an observable given by the tensor product of $N$ Pauli operators, which has seminorm ${3}^{N/2}$. However, the seminorm is a property of the observable and not of the dimension of the space: an observable $O$ and its extension to a larger space, corresponding to the same physical observable and obtained as the tensor product of $O$ with the identity in the additional dimensions, have the same seminorm (\ref{norm N qubits}). Irrespective of $N$, because of Eq. (\ref{variance}) and (\ref{general variance}), the expectation values of all observables with unit seminorm (\ref{norm N qubits}) can be obtained with statistical error bounded by $1/\sqrt{M}$. Any observable of ${\cal B}(\cal H)$ with seminorm different from zero can be obtained from an observable with unit seminorm through multiplication by a real number, representing in fact the observable's seminorm. For a given number of identical preparations and measurements, the statistical error in the expectation value of the generic observable has a bound only dependent on how ``large'' the observable is. In Appendix \ref{app N qubits}, we also explain how the square of the seminorm ${\Vert O\Vert}_2$ defined by
\begin{align}
{\Vert O\Vert}_2^2=\sum_{i\neq 0_v}3^{r_{i}}\,a_i^2, 						\label{original norm}
\end{align}
with $r_i$ representing the power of the monomial $i$ of Pauli operators, namely the number of indices $i_k\neq0$ in $i$, is generally a good approximation of an upper bound of $\var(R[O])$. This result is important for the applications of Sections \ref{The tomographic optimiser} and \ref{Conclusions and outlook}, since ${\Vert O\Vert}_2$ can be significantly smaller than $\Vert O\Vert$ and since, for a chosen statistical error, the number of identical preparations required is directly proportional to $\var(R[O])$.

\par
\vskip 1em

With Eq. (\ref{general estimator}), (\ref{general variance}) and (\ref{norm N qubits}) we have all the elements needed to define the tomographic procedure for a system of $N$ qubits: sample $(\theta_1,\varphi_1,\ldots,\theta_N,\varphi_N)$ uniformly from the surfaces of $N$ unit spheres, rotate each qubit with $e^{i\vartheta_k\vec s\cdot\vec n_{k_{\perp}}}$, measure in the computational basis, obtaining $(m_1,\ldots,m_N)$, calculate ${R}[O](m_1,{\vec n}_1,\ldots,m_N,{\vec n}_N)$ using (\ref{general estimator}), (\ref{identity estimator}) and (\ref{pauli estimator}). Repeat $M$ times and compute $O_M$ of (\ref{monte carlo}), which is equal to $\langle O \rangle$ with normal statistical error, for $M$ sufficiently large, smaller than or equal to ${\Vert O\Vert}/{\sqrt{M}}$. Apart from preparation and measurement, each step of the tomographic procedure only consists in a single--qubit rotation applied to each qubit. As a result, the procedure is easy to implement on a quantum computer simulator and, we believe, is suitable for NISQ computers. As a final remark, we observe that the role of the quantum computer in the procedure is limited to what quantum computers should excel at: generating efficiently probability distributions, in this case $p(m_{1},{\vec n}_1,\ldots,m_{N},{\vec n}_N)=p(m_{s_1},{\vec n}_1,\ldots,m_{s_N},{\vec n}_N)$, difficult to access classically.

%%%%%%%%%%%%%%%%%%%%%%%%%%%%%%%%%%%%%%%%%%%%%%
%                            				THE APPROXIMATE QUANTUM STATE                                         %
%%%%%%%%%%%%%%%%%%%%%%%%%%%%%%%%%%%%%%%%%%%%%%

%\maketitle
\section{The approximate quantum state} \label{approximate description}

The tomographic procedure described consists in the calculation of ${R}[O](m_1,{\vec n}_1,\ldots,m_N,{\vec n}_N)$ after each vector $(m_1,{\vec n}_1,\ldots,m_N,{\vec n}_N)$ is obtained. However, a set of vectors $(m_1,{\vec n}_1,\ldots,m_N,{\vec n}_N)$ 
can be obtained and stored independently of the calculation of a specific estimator and can subsequently be used to calculate the approximate expectation value of any given observable. This is formalised as follows. We call a snapshot $s_j=({m_1}_j,{{}{\vec n}_1}_j,\ldots,{m_N}_j,{{}{\vec n}_N}_j)$ of a system of $N$ qubits the vector with components given by one set of sampled angles and measured values, obtained with one execution of the tomographic procedure described in the previous section. A set 
$\cal S$ of $M$ independent snapshots of the system of $N$ qubits, with the rule 
\begin{align}
{\langle O\rangle}_{{\cal S}}=\frac 1M\sum_{j=1}^M{R}[O](s_j)	\label{approximate expectation values}
\end{align}
for the calculation of the expectation value of a generic observable $O$ as in (\ref{general O}) with $R[O]$ as in (\ref{general estimator}), represents an approximate description of the quantum state of the system of $N$ qubits, as it allows to approximately calculate the expectation value of any observable. The approximation has normal statistical error, bounded by ${\Vert O\Vert}/{\sqrt{M}}$, for $M$ sufficiently large and is independent of $N$.

Going forward, we will refer to $\cal S$ as an approximate quantum state of the system of $N$ qubits. The dimension of the elements of $\cal S$, namely the snapshots, grows linearly in $N$, but the approximation in the computation of the expectation value of the generic observable $O$ only depends on $O$ and the cardinality $M$ of $\cal S$. An approximate quantum state, therefore, represents an approximate, but alternative, description of a quantum system, one whose determination requires a number of copies of the system independent of $N$ and whose storage requires resources only growing linearly in $N$. Clearly, for each quantum state, there are infinitely many approximate quantum states of given cardinality, which are, however, all equivalent for the calculation of the expectation values of observables with the rule (\ref{approximate expectation values}). Even if derived here from density operators and Born's rule, approximate quantum states with the rule (\ref{approximate expectation values}) provide a description of a physical system in terms of statistical inferences based on observed experimental data and, in principle, could be assumed as primitive concepts instead, or, at least, as not requiring the existence of density operators and Born's rule. 

\par
\vskip 1em

To preserve the operator nature of density operators, we could have defined the approximate quantum state as $\tilde{\rho}=\frac 1M\sum_{j=1}^M{K(s_j)}$ and evaluated its distance in ${\cal B(\cal H)}$, defined for example as trace distance from the density operator of the system, as a measure of the quality of the approximation. However, apart from the return to an inconvenient description that is exponentially large in $N$, the distance between operators does not only depend on the operators, it also depends on the definition of distance and, for the evaluation of the quality of the approximation, the choice is quite arbitrary: after all, the error in the calculation of physical quantities, namely expectation values of observables, is the measure of the quality of the approximation that is physically relevant and, for such calculations, $\tilde{\rho}$ would lead again to (\ref{approximate expectation values}). Differently from ${\cal S}$, however, $\tilde{\rho}$, because of its exponential number of matrix elements in $N$, would introduce exponential errors when the estimators of (\ref{approximate expectation values}) are calculated by explicitly multiplying $\tilde{\rho}$ with the observable and taking the trace of the product. We may still define concepts analogous to trace distance or fidelity for approximate quantum states, but instead of functions of operators or distances in ${\cal B(\cal H)}$, we could refer to averages of absolute differences in expectation values of observables. For example, we could define the fidelity between a quantum state and an approximate quantum state as the average distance between exact expectation values of observables and the values of (\ref{approximate expectation values}), which, in the absence of errors in the construction of $\cal S$ and for the same physical state, is, up to a constant factor, bounded by $1/{\sqrt{M}}$. The fidelity between two approximate quantum states could similarly be formalised as the average distance over observables between the values of (\ref{approximate expectation values}) for the two approximate quantum states. 

\par
\vskip 1em

Before we consider possible applications resulting from the availability of an approximate description of a quantum system larger than just a few tens of qubits, we will examine a bit more carefully why it is possible to define ${\cal S}$ with rule (\ref{approximate expectation values}) applicable to any observable. In particular, in the next section, we will consider whether the choice ${\cal G}=SU(2)^{N}$ has an essential role in the definition of the approximate quantum state.

%%%%%%%%%%%%%%%%%%%%%%%%%%%%%%%%%%%%%%%%%%%%%%
%                                                        IS SU(2) SPECIAL?                                                         %
%%%%%%%%%%%%%%%%%%%%%%%%%%%%%%%%%%%%%%%%%%%%%%

%\maketitle
\section{Is \boldmath{$SU(2)$} special?} \label{SU(2) special}

We will start by responding to the question directly: no, $SU(2)$ is not special, it is only one possible convenient choice when observables are expressed as in (\ref{general O}). To explain this, let us consider a tomographic procedure based on the more standard measurements of Pauli operators along the $x,y$ and $z$ directions only. We show in Appendix \ref{pauli} that a procedure equivalent to the one introduced for $SU(2)^{N}$, but with measurements limited to Pauli operators along the $x,y$ and $z$ directions, can be defined with Eq. (\ref{expectation values}), for an observable $O$ as in (\ref{general O}), given by
\begin{align}
{\langle O\rangle}=&\bigg(\frac{1}{4^N}\prod_{k=1}^N \,\sum_{i_k=0}^{3}\,\sum_{m_{i_k}}\bigg)\,p(m_{i_1},\ldots,m_{i_N})\nonumber\\
&\times R[O](m_{i_1},\ldots,m_{i_N}),				\label{tomo N-qubit}
\end{align}
with the estimator $R[O]=4^N\,a_i\,m_{i_1}\cdots m_{i_N}$ and $p(m_{i_1},\ldots,m_{i_N})$ probability of obtaining $(m_{i_1},\ldots,m_{i_N})$ measuring $\sigma_{i_1},\ldots,\sigma_{i_N}$.
In this case, we would sample $(i_1,\ldots,i_N)$ uniformly from $\{0,1,2,3\}^{N}$ and we would measure $\sigma_{i_1},\ldots,\sigma_{i_N}$ (the identity only has one distinct eigenvalue and does not need to be measured: if $i_k=0$ is sampled, we can directly assign 1 to $m_{i_k}$ in the estimator).
If we consider an observable $O$, which only has a limited number, for example polynomial in $N$, of coefficients $a_i$ different from zero, then, given the samples are taken from a set of exponential cardinality in $N$, the probability that $R[O]=0$ for each sample is extremely high, making the Monte Carlo technique ineffective and implying that the tomographic procedure based on the measurement of $\sigma_x,\,\sigma_y,\,\sigma_z$ cannot be used to create an approximate quantum state, since it would not allow to calculate the expectation value of a generic observable as in (\ref{approximate expectation values}). The reason for the high probability of generating samples that are not useful for Monte Carlo does not come from the tomographic procedure alone, but from its combination with the assumed expansion (\ref{general O}) of $O$. In fact, the tomographic procedure based on the measurement of $\sigma_x,\,\sigma_y$ and $\sigma_z$ is derived from the expansion of the density operator on the same orthogonal basis $\{\sigma_{i_1}\cdots\sigma_{i_N}\}$ in $\cal B(\cal H)$ used for the observables in (\ref{general O}) and, at most, one addend of (\ref{general O}) has a projection different from zero on a vector sampled from the orthogonal basis $\{\sigma_{i_1}\cdots\sigma_{i_N}\}$. Instead, the complete set $\{\bigotimes_{k=1}^N e^{i{\psi}_k{\vec s}_k\cdot{\vec n}_k}\}$ of $\cal B(\cal H)$, used for the expansion of the density operator in $SU(2)$ tomography, is such that all components of $O$ in (\ref{general O}) have a projection different from zero on vectors sampled from it. If we inverted the role of the two complete sets, we could use the tomographic procedure based on the measurement of $\sigma_x,\,\sigma_y$ and $\sigma_z$, by expanding the observables as in (\ref{general tomo}), with ${\cal G}=SU(2)^{N}$ and $T(g)=\bigotimes_{k=1}^N e^{-i{\psi}_k{\vec s}_k\cdot{\vec n}_k}$, instead. This would lead to results formally similar to the ones described for $SU(2)$ tomography and $O$ as in (\ref{general O}), with the seminorm (\ref{norm N qubits}) assuming an integral form. However, considering observables of practical interest are typically written as in (\ref{general O}) and not as linear combinations of operators of the form $\bigotimes_{k=1}^N e^{i{\psi}_k{\vec s}_k\cdot{\vec n}_k}$, the $SU(2)$ tomographic procedure turns out to be practically more convenient than a procedure based on Pauli measurements.

\par
\vskip 1em

We finally observe that (\ref{tomo N-qubit}) may not be usable for a generic observable $O$, but can be used with a priori knowledge of $O$. We show in Appendix \ref{pauli} how we can modify, in fact, the domain and the estimator of (\ref{tomo N-qubit}) to include only the points in $\{0,1,2,3\}^{N}$ corresponding to the coefficients $a_i$ different from zero. The limitation of this approach in a variational context is the need for a new set of measurements for every variation, since the coefficients $a_i$ that are different from zero could change at every variation. This would not happen in the $SU(2)$ tomographic schema, as the measurements are only needed to build $\cal S$ and the expectation value of any variation of an observable can be calculated with (\ref{approximate expectation values}). We will develop this aspect of the approximate quantum state in the next section, as we turn our attention to an application of the framework developed so far.

%%%%%%%%%%%%%%%%%%%%%%%%%%%%%%%%%%%%%%%%%%%%%%
%                                                TOMOGRAPHIC OPTIMISER                                                     %
%%%%%%%%%%%%%%%%%%%%%%%%%%%%%%%%%%%%%%%%%%%%%%
%\maketitle
\section{The approximate quantum state variational optimiser} \label{The tomographic optimiser}

We consider the problem of finding the smallest eigenvalue $o$ of an observable $O$, which could represent for instance the Hamiltonian of a quantum simulation problem or the objective function of a combinatorial optimisation. The eigenvalue $o$ can be found as the result of the optimisation
\begin{align}
o=\min_{\psi\in{\cal H}_{\cal N}}{\langle\psi|O|\psi\rangle}, 				\label{optimisation}
\end{align}
with ${\cal H}_{\cal N}$ containing all vectors of $\cal H$ with unit norm ${\Vert \psi \Vert}\equiv\sqrt{\langle\psi|\psi\rangle}$. We can consider an arbitrary vector $\psi_0$ of ${\cal H}_{\cal N}$ and obtain each vector of ${\cal H}_{\cal N}$ with a unitary transformation applied to $\psi_0$. If we parametrise the unitary transformations with the real variables $\xi_1,\xi_2\ldots$, Eq. (\ref{optimisation}) becomes
\begin{align}
o\leqslant\min_{\xi}{\langle\psi_0|U^{\dagger}\,(\xi)O\,U(\xi)|\psi_0\rangle}, 				\label{parametric optimisation}
\end{align}
with $\xi$ indicating the vector of variables $\xi_1,\xi_2\ldots$ and the equality of (\ref{optimisation}) becoming in general an inequality as all values of $\xi$ might not be sufficient to generate all possible unitary transformations. Variational methods such as the Quantum Approximate Optimization Algorithm (QAOA)~\cite{QAOA} and the Variational Quantum Eigensolver (VQE)~\cite{VQE} would start from a specific $\xi$ and execute a set of state preparations, transformations and measurements to estimate the expectation value of $O$ on $|\psi(\xi)\rangle=U(\xi)|\psi_0\rangle$. A new choice for $\xi$ would then be made and the estimation of the expectation value of $O$ on the new $|\psi(\xi)\rangle$ similarly completed. The procedure would be repeated several times with choices of the subsequent values of $\xi$ driven by an optimisation algorithm. The approximate quantum state with the rule (\ref{approximate expectation values}) give us a different approach to (\ref{parametric optimisation}). In fact, instead of writing (\ref{parametric optimisation}) as the minimum of the expectation value of $O$ on $|\psi(\xi)\rangle$, we apply the unitary transformation to $O$ and, with (\ref{approximate expectation values}), obtain
\begin{align}
o\leqslant\min_{\xi}{\langle\psi_0|O(\xi)|\psi_0\rangle}\approx\min_{\xi}{\langle O(\xi) \rangle}_{{\cal S}_0},		\label{approximate state optimisation}
\end{align}
where ${{\cal S}_0}$ represents an approximate quantum state of $|\psi_0\rangle$ and $O(\xi)=U^{\dagger}(\xi)\,O\,U(\xi)$. Equation (\ref{approximate state optimisation}) tells us that we only need to perform one set of preparations, rotations and measurements to build ${{\cal S}_0}$ and, once ${{\cal S}_0}$ is available, the minimisation can be performed as an optimisation in $\xi$ with classical computational resources. The latter is clearly only true if we can explicitly express ${R}[O(\xi)]$ as a function of $\xi$. For any $U(\xi)$ and $O$ as in (\ref{general O}), the estimator ${R}[O(\xi)]$ is given by
\begin{align}
{R}[O(\xi)]=\sum_i \,a_i\,{R}[U^{\dagger}(\xi)\,\sigma_{i_1}\cdots\sigma_{i_N}\,U(\xi)]  	\label{estimator in xi}
\end{align}
and expressing the functional dependence of ${R}[O(\xi)]$ in $\xi$ is in general difficult for arbitrary unitary transformations $U(\xi)$.

We will start to examine (\ref{estimator in xi}) for a parametric transformation given by the product of generic single--qubit unitaries:
\begin{align}
U'(\zeta)=\bigotimes_{k=1}^N U_1({\zeta}_k),  	\label{U as single unitaries}
\end{align}
where $U_1({\zeta}_k)=e^{-i{\psi}_k\vec s\cdot{\vec n}_k}$, with the angles and versors defined as in \ref{tomo 1 qubit}. Substituting Eq. (\ref{U as single unitaries}) in (\ref{estimator in xi}) gives
\begin{align}
{R}[O(\zeta)]=\sum_i \,a_i\prod_{k=1}^N{R}_1[U_1^{\dagger}(\zeta_k)\,\sigma_{i_k}\,U_1(\zeta_k)].  	\label{1 qubit xi estimator}
\end{align}
If $i_k=0$, ${R}_1[U_1^{\dagger}(\zeta_k)\,\sigma_{i_k}\,U_1(\zeta_k)]=1$ because of Eq. (\ref{identity estimator}). For $i_k\neq 0$, since a rotation applied to a Pauli operator gives a Pauli operator in a different direction, ${R}_1[U_1^{\dagger}(\zeta_k)\,\sigma_{i_k}\,U_1(\zeta_k)]$ can be written as ${R}_1[\sigma\cdot{\vec n}(i_k,{\zeta}_k)]$, where ${\vec n}(i_k,{\zeta}_k)$ represents the direction of the Pauli operator after the rotation. Formula (\ref{pauli estimator}) can be employed to express ${R}_1[\sigma\cdot{\vec n}(i_k,{\zeta}_k)]$ as a function of $\zeta_k$, giving with Eq. (\ref{1 qubit xi estimator}) the functional dependence of ${R}[O(\zeta)]$ in $\zeta$.

The transformation (\ref{U as single unitaries}) does not change the entanglement class of $|\psi_0\rangle$. Given we have no general reason to assume that $|\psi_0\rangle$ is in the same entanglement class as the eigenvector of $O$ associated to $o$, we need to introduce transformations involving more than one qubit at a time. We consider $U''(\eta)$ given by
\begin{align}
U''(\eta)&=U_{2}({\eta}_{\,1,2})\,U_{2}({\eta}_{\,3,4})\cdots U_{2}({\eta}_{\,(N-1),N})\nonumber\\
&=\bigotimes_{k=1}^{N/2} U_{2}({\eta}_{\,(2k-1),2k}),  	\label{U as double unitaries}
\end{align}
with $U_{2}(\eta_{i,j})$ acting on qubits $i$ and $j$ and ${\eta}_{i,j}$ representing a set of real variables. For simplicity, we have  assumed that $N$ is even. If $N$ was odd, we would just leave the last qubit out of the transformation (\ref{U as double unitaries}). The choice of the same 2--qubit parametric transformation $U_2$ for each pair of qubits is not necessary and only the most natural when not making assumptions on any known symmetry of the problem. The functional dependence of ${R}[O(\eta)]$ in $\eta$ obviously depends on the choice of $U_2$, which should guarantee sufficient exploration of ${\cal H}_{\cal N}$, without making the optimisation excessively complex. As an example, we consider the case of a parametric $CNOT$ transformation in the Supplementary Material.

With the transformations (\ref{U as single unitaries}) and (\ref{U as double unitaries}) and with Eq. (\ref{approximate state optimisation}), we can now establish a possible optimisation procedure. We start with the execution of the tomographic procedure to construct ${\cal S}_0$. We define $U(\xi)=U'(\zeta)U''(\eta)$ and calculate $o_0=\min_{\xi}{\langle O(\xi) \rangle}_{{\cal S}_0}$. The operators $U''$ drive the optimisation with respect to the entanglement class and the operators $U'$ the optimisation within the entanglement class. We indicate the value of $\xi$ determined through the optimisation and corresponding to the minimum of ${\langle O(\xi) \rangle}_{{\cal S}_0}$ as $\xi^{(0)}=(\zeta^{(0)},\eta^{(0)})$. We prepare $|\psi_1\rangle\equiv U(\xi^{(0)})|\psi_0\rangle$, starting from $|\psi_0\rangle$ and applying $U(\xi^{(0)})$, and build the approximate quantum state ${\cal S}_1$ of $|\psi_1\rangle$ with the tomographic procedure. We use again $U(\xi)=U'(\zeta)U''(\eta)$ and determine $o_1=\min_{\xi}{\langle O(\xi) \rangle}_{{\cal S}_1}$, but we change the operators $U_2$ of $U''$ to act on different qubits pairs, for example $\{1,3\},\{2,4\},\ldots,\{N-2,N\}$. The optimisation will determine a point $\xi^{(1)}$, which can be used to prepare $|\psi_2\rangle=U(\xi^{(1)})U(\xi^{(0)})|\psi_0\rangle$, which will be approximated with ${\cal S}_2$, to calculate $o_2$. The procedure can continue with more iterations, for as long as the depth of $U(\xi^{(t-1)})\cdots U(\xi^{(0)})$, with $t$ indicating the number of iterations, continues to correspond to negligible effects of gates errors.
The final minimum $o_t$ is the approximation of $o$. This procedure is in part similar to the preparation of the hardware-efficient trial states of~\cite{kandala} with one important difference: the entanglers of the hardware-efficient procedure are not necessarily getting the variational state closer to the entanglement class of the lowest eigenstate of $O$. For example, if the lowest eigenstate of $O$ was simply a product state in the computational basis and $|\psi_0\rangle=|0\rangle\cdots|0\rangle$, the application of generic multi--qubit gates would only distance the variational state from the solution. The use of parametric multi--qubit gates as $U''(\eta)$, which need be chosen to satisfy $\exists \,\overline{\eta}: U''(\overline{\eta})=\mathbb{1}$, ensures that $o_t\leqslant o_{t-1}\leqslant \ldots \leqslant o_0$, given unnecessary $U'$ and $U''$ transformations will at worst leave the variational state unchanged, since the optimisation should drive the variables towards values $(\overline{\zeta},\overline{\eta})$ corresponding to $U'(\overline{\zeta})=\mathbb{1} \vee U''(\overline{\eta})=\mathbb{1}$.

\par
\vskip 1em

The approximation of (\ref{approximate state optimisation}) depends on the cardinality of the quantum approximate state, but also on the seminorm of the observable. If $o_t$ is used to approximate $o$, given $o_t=\min_{\xi}{\langle O(\xi) \rangle}_{{\cal S}_t}$, then we should examine how the seminorm of $O(\xi)$ changes as a function of $\xi$. This can actually be avoided by introducing a final step in the optimisation procedure: we can prepare $|\psi_{t+1}\rangle=U(\xi^{(t)})\cdots U(\xi^{(0)})|\psi_0\rangle$, construct ${{\cal S}_{t+1}}$ and calculate and use ${\langle O \rangle}_{{\cal S}_{t+1}}$ as the approximation of $o$, with statistical error bounded by $\Vert O \Vert$ divided by the square root of the cardinality of ${\cal S}_{t+1}$, independent of $\xi$. This final step gives us control of the final statistical error, but the effects of the dependence on $\xi$ of $\Vert O(\xi) \Vert$ are still present at each step of the optimisation procedure and could lead to suboptimal $\xi^{(0)}\ldots\xi^{(t)}$, making the evaluation of how the seminorm of $O(\xi)$ changes as a function of $\xi$ important. We show in Appendix \ref{norm changes} that for $U'(\zeta)$ of (\ref{U as single unitaries}) ${\Vert O \Vert}_2={\Vert O(\zeta) \Vert}_2, \forall{\zeta}$, but also that, differently from the case of single--qubit transformations, ${\Vert O \Vert}_2$ is not maintained in the transformation $U''(\eta)$, namely ${\Vert O \Vert}_2\neq{\Vert O(\eta) \Vert}_2$. We derive the upper bound
\begin{align}
{\Vert O(\eta) \Vert}_2 \leqslant 3^{P/2} {\Vert O \Vert}_2,  	\label{bound in P}
\end{align}
with $P=\min\{Q,N/2\}$, where $Q$ is the highest power of the Pauli monomials of $O$. Equation (\ref{bound in P}) gives us the approximate requirement for the number of additional system's preparations to guarantee a chosen precision in the search of the minimum. The requirement is approximate as we have chosen to use the seminorm (\ref{original norm}) instead of (\ref{norm N qubits}). The choice is motivated by the fact that, as shown in Appendix \ref{app N qubits}, (\ref{original norm}) is generally a good approximation of (\ref{norm N qubits}) and that a bound with (\ref{norm N qubits}) instead would lead to an excessive requirement for the number of preparations. As explained, the final statistical error is linked to the estimation of ${\langle\psi_{t+1}|O|\psi_{t+1}\rangle}$ with ${\langle O \rangle}_{{\cal S}_{t+1}}$ and is, therefore, not affected by this choice.

The presence of powers of 3 in (\ref{norm N qubits}) and in (\ref{bound in P}) appears to be the main limitation in the optimisation procedure, as the possibility of maintaining an established precision by increasing $M$ could become unfeasible as $N$ is increased. In fact, the statistical error of the approximation (\ref{approximate state optimisation}), according to (\ref{norm N qubits}) and (\ref{norm 1 inequality}), is bounded by $3^{Q/2}\sum_{i\neq 0_v}|a_i|$ and, depending on how $Q$ grows with $N$, could be exponential in $N$. Similarly, the additional statistical error of (\ref{bound in P}) could grow rapidly in $N$, depending on how $P$ varies as a function of $N$. The functions $Q=Q(N)$ and $P=P(N)$ clearly depend on the choice of the problem and a few considerations on problems of practical interest seem relevant at this point. In a quantum simulation context, if $O$ represents the electronic Hamiltonian with Bravyi-Kitaev mapping to Pauli operators~\cite{BK},  $Q(N)$ is $O(\log N)$ and the factor $3^{Q/2}$ becomes $O(N^{c(\log 3)/2})$, with $c$ real number dependent on the system. If we also include the statistical errors coming from (\ref{bound in P}), then, since for $N$ large enough $P=Q$, we obtain again a statistical error with bound $O(N^{c(\log 3)/2})$, for an overall statistical error of the optimisation procedure growing at most polynomially in $N$. For some cases, with problem--specific mappings to Pauli operators, we can improve on the dependence of $P$ and $Q$ on $N$ or even eliminate it altogether, as for example in~\cite{hubbard}, where, even if at the expense of using ancillary qubits, the 2D Hubbard model is mapped to a 4-local qubit Hamiltonian for any $N$. Combinatorial optimisations are also of the form (\ref{optimisation}), possibly with additional constraints of the form $\psi\in{\cal H'}\subset{\cal H}_{\cal N}$, and have constant $P$ and $Q$ in $N$. For all common instances with $P=Q=2$, the factors $3^{Q/2}$ and $3^{P/2}$ contribute to the number of additional system's preparations to maintain a given precision only with a factor smaller than ten.

\par
\vskip 1em

The optimisation procedure described is a specific realisation of Eq. (\ref{approximate state optimisation}). Clearly, the choices made for $U(\xi)$ and for the decomposition in $t$ iterations are far from unique. Explicit verifications and considerations on the simplicity of the optimisation, the quantum circuit's complexity and the depth in the exploration of ${\cal H}_{\cal N}$ could be made for a variety of cases. Furthermore, some problems might have additional requirements, adding to the factors for consideration. For example, in the presence of hard constraints in a combinatorial optimisation, we could utilise the same methods of~\cite{QAOAnew}, introduced for QAOA, creating requirements on the form of $U(\xi)$, in order to satisfy the constraints. Similarly, $U(\xi)$ should be limited to conservative gates for a quantum simulation problem with constant particle number. Still and all, for any choice of $U(\xi)$ and of the decomposition in $t$ iterations, the defining and valuable feature of the approximate quantum state optimiser is its access to the explicit functional dependence of the objective function on the optimisation variables, which produces a significant reduction in the total number of iterations with respect to variational methods, which, in order to evaluate the objective function, require multiple and independent iterations for every value of the optimisation variables considered. Finally, we note that an optimisation procedure based on Eq. (\ref{approximate state optimisation}) is also well--suited for parametric objective functions, as, for example, for the case of an electronic Hamiltonian with parametric internuclear distances or for a mixed binary optimisation\cite{braine}. In fact, the only difference in the procedure would be represented by the execution of the classical optimisations on both the optimisation variables and the objective function's parameters.

%%%%%%%%%%%%%%%%%%%%%%%%%%%%%%%%%%%%%%%%%%%%%%
%                                                CONCLUSIONS AND OUTLOOK                                                  %
%%%%%%%%%%%%%%%%%%%%%%%%%%%%%%%%%%%%%%%%%%%%%%
%\maketitle
\section{Conclusions and outlook} \label{Conclusions and outlook}

We have introduced the concept of approximate quantum state for a system of $N$ qubits and have shown that the dimension of the space of the system does not influence the number of copies of the system we need, in order to gain information on measurable quantities. The proof we have used is constructive: the approximate quantum state comes with an explicit operational procedure to estimate the expectation value of a generic observable with statistical error only dependent on the cardinality of the approximate quantum state and on an observable's seminorm, independent of $N$. The operational procedure, other than the initial state preparations and final state measurements, only requires single--qubit rotations, which suggests that constructing approximate quantum states could be appropriate for NISQ computers. The implementation of the operational procedure on quantum hardware and the observation of the effects of noise on the approximate quantum state and on the estimation of the observables' expectation values represent important, immediate next steps. The definition of the approximate quantum state enabled the introduction of a variational optimisation method, which does not require separate iterations for each set of values of the variables. The definition of the optimal unitaries and steps appearing in the variational optimisation method will benefit from further theoretical and numerical work. This could include examining the combination of the variational optimisation method with other known variational methods, such as VQE, with the former determining the initial state then used by the latter in a final optimisation cycle.

\par
\vskip 0.96em

The variational optimisation method introduced, which can be applied to various problems, such as quantum simulations and combinatorial optimisations, is easily generalised by considering optimisations different from (\ref{optimisation}). Machine learning optimisations, in particular for generative models known as Born Machines~\cite{Bornoriginal}, are a significant and interesting case for the approximate quantum state optimiser. In fact, the explicit dependence of the expectation values of the projectors for the bitstrings on the variational parameters~\cite{Bornmarcello} can be obtained with (\ref{approximate expectation values}) even for a large number of qubits, since, as shown in Appendix \ref{app N qubits}, the estimators for the projectors for the bitstrings have, with high probability, variance bounded by 1 for all values of $N$.

\par
\vskip 0.96em

The definition of the snapshots, constituting an approximate quantum state, can be generalised, too. The snapshots can in fact be derived from a group different from $SU(2)$. This generalisation has interest that goes beyond the purely mathematical exercise. As we have seen, $SU(2)$ tomography works well for observables expressed as linear combinations of products of Pauli operators, especially if the powers of the Pauli monomials are low. Electronic Hamiltonians can be represented as linear combinations of products of Pauli operators, but at the cost, in general, of introducing Pauli monomials with powers increasing with $N$. In second quantisation, electronic Hamiltonians contain linear combinations of products of only two or four ladder operators for any $N$. The bosonic tomographic formulas of~\cite{Paini2000Quantum} can be extended to the fermionic case, choosing the additive group of Grassmann numbers~\cite{fermions}, instead of the additive group of complex numbers, as tomographic group and expressing the snapshots as vectors of sampled real numbers descending from Grassmann numbers and of measured presence or absence of electrons in orbitals. The estimators for ladder operators and for their linear combinations can then be calculated directly without the need for the mapping to Pauli operators and without the consequent appearance of powers of Pauli monomials growing with $N$.

\par
\vskip 0.96em

The methods of fermionic tomography and generative modelling utilising the approximate quantum state are being actively developed by the authors.

%%%%%%%%%%%%%%%%%%%%%%%%%%%%%%%%%%%%%%%%%%%%%%
%                                                      ACKNOWLEDGEMENTS                                                  %
%%%%%%%%%%%%%%%%%%%%%%%%%%%%%%%%%%%%%%%%%%%%%%

\section{Acknowledgments}

MP would like to thank Chad Rigetti and Michael Brett for the support, Dan Padilha for the numerical simulations, Amy Brown, Colm Ryan, David Garvin, Duncan Fletcher, Eric Peterson, Genya Crossman, John Lapeyre, John Macaulay, Juan Bello-Rivas, Mark Hodson, Max Henderson, Riccardo Manenti, Robert Smith and Sohaib Alam for the reviews and the feedback.

AK acknowledges the support from the US Department of Defense and the support from the AFOSR MURI project ``Scalable Certification of Quantum Computing Devices and Networks''.

%%%%%%%%%%%%%%%%%%%%%%%%%%%%%%%%%%%%%%%%%%%%
%MP: For Sienna, Mila Stella and Billur (and for their patience when I spent weekends doing this). %
%%%%%%%%%%%%%%%%%%%%%%%%%%%%%%%%%%%%%%%%%%%%

%%%%%%%%%%%%%%%%%%%%%%%%%%%%%%%%%%%%%%%%%%%%%%
%                                                           BIBLIOGRAPHY                                                           %
%%%%%%%%%%%%%%%%%%%%%%%%%%%%%%%%%%%%%%%%%%%%%%

\bibliographystyle{apsrev}
\bibliography{biblio}

%%%%%%%%%%%%%%%%%%%%%%%%%%%%%%%%%%%%%%%%%%%%%%
%                                                           APPENDIX 1                                                                %
%%%%%%%%%%%%%%%%%%%%%%%%%%%%%%%%%%%%%%%%%%%%%%
%\maketitle
\begin{widetext}
\section{Appendices}

\subsection{One qubit} \label{app 1 qubit}
\noindent

We start from Eq. (\ref{kernel}) derived in~\cite{Paini2000Quantum,DAriano2003Spin}:
\begin{equation}
K_1(m_s,\vec n)=\frac{2}{\pi}\int_0^{2\pi}d\psi\sin^2\frac\psi2 e^{i\psi(m_s-\vec s\cdot\vec n)},  \label{k1 again}
\end{equation}
where $m_s$ is the result of the measurement of the spin operator $\vec{s}$ along a unit vector $\vec{n}=(\cos \varphi\sin\vartheta,\sin\varphi\sin\vartheta,\cos\vartheta)$, with $(\vartheta,\varphi)\in[0,\pi]{\times}[0,2\pi)$.  We  easily rewrite (\ref{k1 again}) with Pauli operators $\vec{\sigma}=(\sigma_x,\sigma_y,\sigma_z)=2\,\vec{s}$ and their eigenvalues $m=2\,m_s=\pm1$ instead of spin operators as
\begin{equation}
K_1(m,\vec{n})=\frac{2}{\pi}\int_0^{2\pi}d\psi\sin^2\frac\psi2 e^{i\frac{\psi}{2}(m-\vec\sigma\cdot\vec n)}. \label{k1 with pauli}
\end{equation}
We calculate (\ref{k1 with pauli}) utilising the identity $e^{- i\frac{\psi}{2}\vec{\sigma}\cdot\vec{n}}=\mathbb{1}\cos{\frac{\psi}{2}}- i\,\vec{\sigma}\cdot\vec{n}\,\sin{\frac{\psi}{2}}$ and obtain
\begin{equation}\label{eq:k1}
K_1(m,\vec{n})=\frac{1}{2}(\mathbb{1}+ 3\,m\,\vec{\sigma}\cdot\vec{n}).
\end{equation}
The 1--qubit estimator for an observable $O$ is given by $R_1[O](m,\vec{n})=\tr[O K_1(m,\vec{n})]$ and, with Eq.~\eqref{eq:k1}, becomes
\begin{equation}
R_1[O](m,\vec{n})=\frac{1}{2}\big(\tr[O]+ 3\,m\,\tr[O\,\vec{\sigma}]\cdot\vec{n}\big).  \label{estimator1}
\end{equation}
For $O$ equal to the identity, (\ref{estimator1}) is easily calculated:
\begin{equation}
R_1[\mathbb{1}](m,\vec{n})=1,
\end{equation}
as expected in (\ref{identity estimator}). For the case of $O$ equal to a Pauli operator along a unit direction $\vec{u}$, (\ref{estimator1}) gives
\begin{equation}
R_1[\vec{\sigma}\cdot\vec{u}](m,\vec{n})= 3\,m\,\vec{u}\cdot\vec{n},
\end{equation}
which, for $\vec{u}$ in the $x,y$ or $z$ direction, is Eq. (\ref{pauli estimator}):
\begin{equation}
R_1[\sigma_\alpha](m,\vec{n})=3\,m\,n_\alpha, \label{R1 for pauli}
\end{equation}
with $\alpha=x,y,z$ and $n_\alpha$ defined by $\vec{n}=(n_x,n_y,n_z)$. The variance of the estimator $R_1[\sigma_\alpha](m,\vec{n})$ is
\begin{align}
\var(R_1[\sigma_\alpha](m,\vec{n}))=\langle {R}_1^2[\sigma_\alpha](m,\vec n) \rangle-\langle {R}_1[\sigma_\alpha](m,\vec n) \rangle^2\leqslant \langle {R}_1^2[\sigma_\alpha](m,\vec n) \rangle. \label{variance defined}
\end{align}
We indicate with $p(m,\vec{n})$ the probability that a measurement of $\vec{\sigma}\cdot\vec{n}$ gives $m$ and calculate the last term of (\ref{variance defined}) utilising (\ref{R1 for pauli})
\begin{align}\label{variance of pauli}
\langle {R}_1[\sigma_\alpha](m,\vec n){R}_1[\sigma_\beta](m,\vec n) \rangle&=\int_{\Sigma}\frac{d\vec n}{4\pi}\sum_{m=\pm1}p(m,\vec{n}){R}_1[\sigma_\alpha](m,\vec{n}){R}_1[\sigma_\beta](m,\vec{n})\\\nonumber
&=9\int_{\Sigma}\frac{d\vec n}{4\pi}\,n_\alpha\,n_\beta \sum_{m=\pm1}p(m,\vec{n})=9\int_{\Sigma}\frac{d\vec n}{4\pi}\,n_\alpha\, n_\beta=3\,\delta_{\alpha\beta},
\end{align}
with $\delta_{\alpha\beta}$ equal to 1 when $\alpha=\beta$, 0 when $\alpha\neq\beta$. Equation (\ref{variance of pauli}), with (\ref{variance defined}), gives (\ref{pauli variance}). Equation (\ref{identity variance}), on the other hand, is the simple consequence of $R_1[\mathbb{1}](m,\vec{n})$ being a constant. To calculate the variance of $R_1[O](m,\vec{n})$, we write $O$ as a linear combination of the identity and Pauli operators as in (\ref{O})
\begin{align}
O=\sum_{i=0}^{3}a_i\,\sigma_i 															
\end{align}
and utilise the results obtained for the estimators for the identity and Pauli operators:
\begin{align}\label{main on variance}
\var(R_1[O](m,\vec{n}))&=\var\bigg(a_o\,R_1[\mathbb{1}](m,\vec{n}) + \sum_{i=1}^{3}a_i\,R_1[\sigma_i](m,\vec{n})\bigg)\\\nonumber
&=a_0^2\,\var(R_1[\mathbb{1}](m,\vec{n}))+\var\bigg(\sum_{i=1}^{3}a_i\,R_1[\sigma_i](m,\vec{n})\bigg)+2\,\cov\bigg(R_1[\mathbb{1}](m,\vec{n}),\sum_{i=1}^{3}a_i\,R_1[\sigma_i](m,\vec{n})\bigg)\\\nonumber
&=\var\bigg(\sum_{i=1}^{3}a_i\,R_1[\sigma_i](m,\vec{n})\bigg)\leqslant\bigg\langle \bigg(\sum_{i=1}^{3}a_i\,R_1[\sigma_i](m,\vec{n})\bigg)^{\!2}\, \bigg\rangle\\\nonumber
&=\sum_{i=1}^{3}a_i^2\,\langle R_1^2[\sigma_i](m,\vec{n}) \rangle + \sum_{i\neq j (\neq 0)}\,a_i\,a_j \langle R_1[\sigma_i](m,\vec{n})R_1[\sigma_j](m,\vec{n})\rangle\\\nonumber
&=3\sum_{i=1}^{3}a_i^2+3 \sum_{i\neq j (\neq 0)}\,a_i\,a_j \,\delta_{ij}=3\sum_{i=1}^{3}a_i^2.
\end{align}
We define a seminorm in ${\cal B}(\cal H)$ as in (\ref{norm 1 qubit})
\begin{align}
\Vert O\Vert^2=3\sum_{i=1}^{3}a_i^2 										
\end{align}
and obtain Eq. (\ref{O variance}) from (\ref{main on variance}):
\begin{align}
\var({R}_1[O](m,\vec n))\leqslant{\Vert O\Vert}^2.
\end{align}

%%%%%%%%%%%%%%%%%%%%%%%%%%%%%%%%%%%%%%%%%%%%%%
%                                                           APPENDIX 2                                                                %
%%%%%%%%%%%%%%%%%%%%%%%%%%%%%%%%%%%%%%%%%%%%%%

\subsection{\boldmath{$N$} qubits} \label{app N qubits}
\noindent

For a system of $N$ qubits, we expand the generic observable $O$ as in (\ref{general O})
\begin{align}
O=\sum_i \,a_i\,\sigma_{i_1}\cdots\sigma_{i_N}										\label{general O again}
\end{align}
and proceed in the calculation of the variance of the estimator for $O$ as in the one qubit case:
\begin{align}\label{main on variance N qubits}
\var(R[O])&=a_0^2\,\var(R[\mathbb{1}])+\var\bigg(\sum_{i\neq 0_v}a_i\,R[\sigma_{i_1}\cdots\sigma_{i_N}]\bigg)+2\,\cov\bigg(R[\mathbb{1}],\sum_{i\neq 0_v}a_i\,R[\sigma_{i_1}\cdots\sigma_{i_N}]\bigg)\\\nonumber
&=\var\bigg(\sum_{i\neq 0_v}a_i\,R[\sigma_{i_1}\cdots\sigma_{i_N}]\bigg)\leqslant\bigg\langle \bigg(\sum_{i\neq 0_v}a_i\,R[\sigma_{i_1}\cdots\sigma_{i_N}]\bigg)^{\!2}\, \bigg\rangle\\\nonumber
&=\sum_{i,j (\neq 0_v)}\,a_i\,a_j \langle R[\sigma_{i_1}\cdots\sigma_{i_N}]R[\sigma_{j_1}\cdots\sigma_{j_N}]\rangle,\nonumber
\end{align}
where the dependence of all estimators on $(m_1,\vec{n}_{1},\ldots,m_N,\vec{n}_{N})$ has not been explicitly indicated for a lighter notation and the last sum on $i$ and $j$ is extended to all values of $i$ and $j$ except $0_v\equiv(0,0,\ldots,0)$. In general, for every $k$, the indices $i_k$ and $j_k$ appearing in each $\langle R[\sigma_{i_1}\cdots\sigma_{i_N}]R[\sigma_{j_1}\cdots\sigma_{j_N}]\rangle$ of (\ref{main on variance N qubits}) can assume any integer value between 0 and 4. For given $i$ and $j$, let $S$ be the set of indices $k$ for which $i_k\neq 0 \wedge  j_k\neq 0$, $s$ the cardinality of $S$ and $\overline S$ the set containing the remaining indices $k$ of ${1,\ldots,N}$ with cardinality ${\bar s}=N-s$. We indicate $\sigma_{i_1}\cdots\sigma_{i_N}$ and $\sigma_{j_1}\cdots\sigma_{j_N}$ with $\sigma_{i}$ and $\sigma_{j}$ and use (\ref{estimator product of single qubits estimators}) for a Pauli monomial
\begin{align}\label{mother of all expansions}
\langle R[\sigma_{i}]R[\sigma_{j}]\rangle
&=\bigg(\prod_{{h}\in{\overline S}}\int_{{\Sigma}_{h}}\frac{d{\vec n}_{h}}{4\pi}\sum_{m_{h}}{R}_1[\sigma_{i_{h}}]{R}_1[\sigma_{j_{h}}]\bigg)\bigg(\prod_{k\in{S}}\int_{{\Sigma}_{k}}\frac{d{\vec n}_{k}}{4\pi}\sum_{m_{k}}{R}_1[\sigma_{i_{k}}]{R}_1[\sigma_{j_{k}}]\bigg)p(m_1,\vec{n}_{1},\ldots,m_N,\vec{n}_{N})\\\nonumber
&=\bigg(\prod_{{h}\in{\overline S}}\int_{{\Sigma}_{h}}\frac{d{\vec n}_{h}}{4\pi}\sum_{m_{h}}{R}_1[\sigma_{i_{h}}]{R}_1[\sigma_{j_{h}}]\bigg)\bigg(9^s\prod_{k\in{S}}\int_{{\Sigma}_{k}}\frac{d{\vec n}_{k}}{4\pi}\,{n}_{i_k}{n}_{j_k}\bigg)\bigg(\prod_{k\in{S}}
\sum_{m_{k}}p(m_1,\vec{n}_{1},\ldots,m_N,\vec{n}_{N})\bigg)\\\nonumber
&=\bigg(\prod_{{h}\in{\overline S}}\int_{{\Sigma}_{h}}\frac{d{\vec n}_{h}}{4\pi}\sum_{m_{h}}{R}_1[\sigma_{i_{h}}]{R}_1[\sigma_{j_{h}}]\bigg)\bigg(3^s \prod_{k\in{S}}\delta_{{i_k}{j_k}}\bigg)p(m_{h_1},\vec{n}_{h_1},\ldots,m_{h_{\bar{s}}},\vec{n}_{h_{\bar{s}}})\\\nonumber
&=\bigg(\prod_{{h}\in{\overline S}}\int_{{\Sigma}_{h}}\frac{d{\vec n}_{h}}{4\pi}\sum_{m_{h}}{R}_1[\sigma_{i_{h}}]{R}_1[\sigma_{j_{h}}]\bigg)\bigg(3^s\prod_{k\in{S}}\delta_{{i_k}{j_k}}\int_{{\Sigma}_{k}}\frac{d{\vec n}_{k}}{4\pi}\sum_{m_{k}}{R}_1[\sigma_{i_{k}}^2]\bigg)p(m_1,\vec{n}_{1},\ldots,m_N,\vec{n}_{N})\\\nonumber
&=3^s \bigg( \prod_{k\in{S}}\delta_{{i_k}{j_k}}\bigg)\bigg(\prod_{{h}\in{\overline S}}\int_{{\Sigma}_{h}}\frac{d{\vec n}_{h}}{4\pi}\sum_{m_{h}}{R}_1[\sigma_{i_{h}}\sigma_{j_{h}}]\bigg)\bigg(\prod_{k\in{S}}\int_{{\Sigma}_{k}}\frac{d{\vec n}_{k}}{4\pi}\sum_{m_{k}}{R}_1[\sigma_{i_{k}}\sigma_{j_{k}}]\bigg)p(m_1,\vec{n}_{1},\ldots,m_N,\vec{n}_{N})\\\nonumber
&=3^s \bigg( \prod_{k\in{S}}\delta_{{i_k}{j_k}}\bigg)\langle R[\sigma_{i}\sigma_{j}]\rangle=3^s \bigg( \prod_{k\in{S}}\delta_{{i_k}{j_k}}\bigg)\tr[\sigma_{i}\sigma_{j}\rho]\leqslant3^s \bigg( \prod_{k\in{S}}\delta_{{i_k}{j_k}}\bigg)\equiv3^{r_{ij}}\Delta_{ij}.\nonumber
\end{align}
If we introduce a seminorm in ${\cal B}({\cal H})$ as in (\ref{norm N qubits})
\begin{align}
\Vert O\Vert^2=\sum_{i,j (\neq 0_v)}3^{r_{ij}}\Delta_{ij}|a_i||a_j|
 										\label{norm N qubits again}
\end{align}
Eq. (\ref{main on variance N qubits}) becomes (\ref{general variance}) 
\begin{align}
\var({R}[O])\leqslant{\Vert O\Vert}^2.						
\end{align}

\par
\vskip 1em

The seminorm (\ref{norm N qubits again}) is a stronger upper bound for the variance of $R[O]$ than the formally simpler seminorm
\begin{align}
{\Vert O\Vert}_1=\sum_{i\neq 0_v}{\sqrt 3}^{r_{i}}|a_i|,
 										\label{norm 1}
\end{align}
with $r_i$ representing the power of the monomial $i$ of Pauli operators, namely the number of indices $i_k\neq0$ in $i$, since
\begin{align}\label{norm 1 inequality}
\Vert O\Vert^2=\sum_{i,j (\neq 0_v)}3^{r_{ij}}\Delta_{ij}|a_i||a_j|\leqslant\sum_{i,j (\neq 0_v)}{\sqrt 3}^{r_{i}}{\sqrt 3}^{r_{j}}|a_i||a_j|=\Bigl(\sum_{i\neq 0_v}{\sqrt 3}^{r_{i}}|a_i|\Bigr)^2={\Vert O\Vert}_1^2
\end{align}
This is not surprising, since the seminorm (\ref{norm 1}) can be shown to be a bound for the variance of $R[O]$ by expanding $\var\bigl(\sum_{i\neq 0_v}a_i\,R[\sigma_{i_1}\cdots\sigma_{i_N}]\bigr)$ of (\ref{main on variance N qubits}) and using Schwarz inequality without resorting to explicit properties of $R$ as in (\ref{mother of all expansions}). The seminorm (\ref{norm N qubits again}) can also be expressed as
\begin{align}
\Vert O\Vert^2={\Vert O\Vert}_2^2+\sum_{i\neq j (\neq 0_v)}3^{r_{ij}}\Delta_{ij}|a_i||a_j|,			\label{norm decomposition}
\end{align}
with the seminorm ${\Vert O\Vert}_2$ defined by
\begin{align}
{\Vert O\Vert}_2^2=\sum_{i\neq 0_v}3^{r_{i}}\,a_i^2. 						\label{original norm N qubits}
\end{align}
We use (\ref{mother of all expansions}) and rewrite (\ref{main on variance N qubits}) as
\begin{align}
\var(R[O])\leqslant\sum_{i,j (\neq 0_v)}\,3^{r_{ij}}\,\Delta_{ij}\,\tr[\sigma_{i}\sigma_{j}\rho]\,a_i\,a_j={\Vert O\Vert}_2^2+\sum_{i\neq j (\neq 0_v)}\,3^{r_{ij}}\,\Delta_{ij}\,\tr[\sigma_{i}\sigma_{j}\rho]\,a_i\,a_j.		\label{variance and mixed terms}
\end{align}
Equation (\ref{variance and mixed terms}) shows how the mixed terms of (\ref{norm decomposition}) are a weak upper bound for the last term of (\ref{variance and mixed terms}). For a randomly chosen $O$ or, in the same way, for a random $\rho$, the addends in the last term of (\ref{variance and mixed terms}) have mixed positive and negative signs. Furthermore, for most states $\rho$, the expectation values of the expressions $\sigma_i \sigma_j$ tend to become smaller for higher values of $r_{ij}$. As a result, ${\Vert O\Vert}_2$ is generally a good approximation of the standard deviation of $R[O]$.

To formalise this statement, we start by showing that the expectation value of the mixed terms of (\ref{variance and mixed terms}) over the space of the density operators of the system is equal to zero. The probability of choosing a density operator from the space of density operators is the probability of choosing a set of eigenvalues multiplied by the probability of choosing a density operator with specific eigenvalues conditional to the choice of the eigenvalues. If for any choice of the eigenvalues, the expectation value over the density operators with the chosen eigenvalues is zero, then the expectation value over all density operators is zero. All density operators with the same eigenvalues can be obtained from one with a unitary transformation and, vice versa, all unitary transformations applied to a density operator preserve the eigenvalues. Therefore, we can calculate the expectation value of the mixed terms of (\ref{variance and mixed terms}) over the density operators with given eigenvalues by averaging over the unitary group~\cite{Li2013SelectedTI} with the unitary transformations applied to an arbitrary density operator $\rho_0$
\begin{align}\label{unitary average}
&\int_U dU\sum_{i\neq j (\neq 0_v)}3^{r_{ij}}\,\Delta_{ij}\,a_i\,a_j\,\tr[\sigma_{i}\sigma_{j}\,U\rho_0\,U^{\dag}]
=\sum_{i\neq j (\neq 0_v)}3^{r_{ij}}\,\Delta_{ij}\,a_i\,a_j\,\tr\bigg[\sigma_{i}\sigma_{j}\int_U dU\, U\rho_0\,U^{\dag}\bigg]\\
&=\sum_{i\neq j (\neq 0_v)}3^{r_{ij}}\,\Delta_{ij}\,a_i\,a_j\,\tr\Bigl[\frac{\sigma_{i}\sigma_{j}}{2^N}\Bigr]=0,\nonumber
\end{align}
which implies
\begin{align}\label{overall average}
\bigg\langle \sum_{i\neq j (\neq 0_v)}\,3^{r_{ij}}\,\Delta_{ij}\,\tr[\sigma_{i}\sigma_{j}\rho]\,a_i\,a_j\bigg\rangle_{\!\!\rho}=0.
\end{align}
To gain further information on how close ${\Vert O\Vert}_2$ is likely to be to ${\Vert O\Vert}$ for a random state, we now consider the variance of the mixed terms of (\ref{norm decomposition}) over the space of the density operators of the system. For simplicity and in line with the needs of the applications of Sections \ref{The tomographic optimiser} and \ref{Conclusions and outlook}, we specialise the analysis to pure states. From the unitary group notation of (\ref{unitary average}), we switch to the average over the states expressed as in~\cite{Ambainis:2007:QTT:1251970.1252159} and recall the following identities:
\begin{align}
\int_\psi{d\psi}\ket{\psi}\!\bra{\psi}^{\otimes 2}=\frac2{2^N(2^N+1)}\,\mathbb{1}_{\rm sym},
\end{align}
with $\mathbb{1}_{\rm sym}$ projector onto the symmetric subspace of ${\cal H}^{\otimes 2}$, and
\begin{align}
\tr[A\otimes B\,\mathbb{1}_{\rm sym}]=\frac1{2}\big(\tr[A]\tr[B]+\tr[AB]\big),
\end{align}
for any operator $A$ and $B$ of ${\cal B}(\cal H)$. The variance of the mixed terms of (\ref{variance and mixed terms}) over the states of the system can be expressed as
\begin{align}\label{variance of mixed terms}
&\var\bigg(\sum_{i\neq j (\neq 0_v)}\,3^{r_{ij}}\,\Delta_{ij}\,\tr\big[\sigma_{i}\sigma_{j}\ket{\psi}\!\bra{\psi}\big]\,a_i\,a_j\bigg)_{\!\psi}=\int_\psi{d\psi}\bigg(\sum_{i\neq j(\neq 0_v)}a_i\,a_j\,3^{r_{ij}}\,\Delta_{ij}\,\tr\big[\sigma_{i}\sigma_{j}\ket{\psi}\!\bra{\psi}\big]\bigg)^{\! 2}\\\nonumber
&=\sum_{i\neq j(\neq 0_v)}\sum_{i'\neq j'(\neq 0_v)}a_i\,a_j\,a_{i'}a_{j'}\,3^{r_{ij}}\,3^{r_{i'j'}}\,\Delta_{ij}\,\Delta_{i'j'}\int_\psi{d\psi}\,\big(\tr\big[\sigma_{i}\sigma_{j}\ket{\psi}\!\bra{\psi}\big]\,\tr\big[\sigma_{i'}\sigma_{j'}\ket{\psi}\!\bra{\psi}\big]\big)\\\nonumber
&=\sum_{i\neq j(\neq 0_v)}\sum_{i'\neq j'(\neq 0_v)}a_i\,a_j\,a_{i'}a_{j'}\,3^{r_{ij}}\,3^{r_{i'j'}}\,\Delta_{ij}\,\Delta_{i'j'}\,\tr\bigg[(\sigma_i\sigma_j)\otimes (\sigma_{i'}\sigma_{j'})\int_\psi{d\psi}\ket{\psi}\!\bra{\psi}^{\otimes 2}\bigg]\\\nonumber
&=\frac2{2^N(2^N+1)}\sum_{i\neq j(\neq 0_v)}\sum_{i'\neq j'(\neq 0_v)}a_i\,a_j\,a_{i'}a_{j'}\,3^{r_{ij}}\,3^{r_{i'j'}}\,\Delta_{ij}\,\Delta_{i'j'}\,\tr\big[(\sigma_i\sigma_j)\otimes (\sigma_{i'}\sigma_{j'})\,\mathbb{1}_{\rm sym}\big]\\\nonumber
&=\frac1{2^N(2^N+1)}\sum_{i\neq j(\neq 0_v)}\sum_{i'\neq j'(\neq 0_v)}a_i\,a_j\,a_{i'}a_{j'}\,3^{r_{ij}}\,3^{r_{i'j'}}\,\Delta_{ij}\,\Delta_{i'j'}\,\tr[\sigma_i\sigma_j\sigma_{i'}\sigma_{j'}]\\\nonumber
&=\frac1{(2^N+1)}\sum_{i\neq j(\neq 0_v)}\sum_{i'\neq j'(\neq 0_v)}a_i\,a_j\,a_{i'}a_{j'}\,3^{r_{ij}}\,3^{r_{i'j'}}\,\Delta_{ij}\,\Delta_{i'j'}\,\overline{\Delta}_{iji'j'},\\\nonumber
\end{align}
with $\overline{\Delta}_{iji'j'}=1$ when $i,j,i',j'$ are such that $\sigma_i\sigma_j\sigma_{i'}\sigma_{j'}=\mathbb{1}$, otherwise $\overline{\Delta}_{iji'j'}=0$. A general upper bound of the last expression is likely to be weak, given its strong dependence on $O$. We will then examine the last expression for an important special case, namely the case of the observables $\{P\}$, projectors on a generic element of the computational basis, namely
\begin{align}
P=\bigotimes_{k=1}^N\,\frac{1}{2} (\mathbb{1}\pm\sigma_{z_k}),    \label{projector}
\end{align}
where the $\pm$ of each factor is defined by the choice of the element of the computational basis. This case is important not only for its practical significance in a machine learning context, as briefly indicated in Section \ref{Conclusions and outlook}, but also because it is representative of the cases with bounded ${\Vert O\Vert}_2$ and diverging ${\Vert O\Vert}$ for increasing $N$,  which makes the determination of whether the statistical error of the estimator can be approximated by ${\Vert O\Vert}_2$ of particular interest. To calculate the seminorms of any $P$, we expand the tensor product of (\ref{projector}) and use the seminorm definitions. For ${\Vert P\Vert}_2$ this simply means
\begin{align}
{\Vert P\Vert}_2^2=\frac{1}{4^N}\bigg({{N}\choose{N-1}}\,3^1+{{N}\choose{N-2}}\,3^2+\ldots+{{N}\choose{0}}\,3^N\bigg)=\frac{1}{4^N}\big((1+3)^N\big)-\frac{1}{4^N}=1-\frac{1}{4^N}.
    \label{norm_2 projector}
\end{align}
A more involved combinatorial calculation would show that ${\Vert P\Vert}^2\leqslant(3/2)^N$ and ${\Vert P\Vert}^2=O((3/2)^N)$. We come back to (\ref{variance of mixed terms}) for any $P$ and observe that, for given $i$ and $j$, there are $2^N$ ordered pairs $i',j'$ such that $\overline{\Delta}_{iji'j'}=1$. The operator $\sigma_i\sigma_j$ is in fact the tensor product of single--qubit identities and operators $\sigma_z$. Each of them becomes the identity when multiplied by itself and this can be obtained with two choices for every qubit of $\sigma_{i'}$ and $\sigma_{j'}$. Unfortunately, to complicate things, ${r_{i'j'}}$ is not independent of the choice of $i'$ and $j'$ for given $i$ and $j$; however, we can simplify the evaluation of (\ref{variance of mixed terms}) by considering a limit case and determining an upper bound. In particular, for every $i$ and $j$, with $i\neq j$ and $i$ and $j$ different from $0_v$,
\begin{align}    \label{variance of mixed terms of projectors}
\sum_{i',j'}\,3^{r_{i'j'}}\overline{\Delta}_{iji'j'}<\sum_{i',j'}\,3^{r_{i'j'}}\overline{\Delta}_{iii'j'}=\sum_{l=0}^N{{N}\choose{l}}\,3^{l}=4^N.
\end{align}
With the substitutions $a_i=(1/2)^N$ and $\Delta_{ij}=1$ valid for any $P$ and any value of the indexes, the last term of Eq. (\ref{variance of mixed terms}) becomes
\begin{align}\label{smaller than 3/4}
&\frac1{(2^N+1)}\sum_{i\neq j(\neq 0_v)}\sum_{i'\neq j'(\neq 0_v)}a_i\,a_j\,a_{i'}a_{j'}\,3^{r_{ij}}\,3^{r_{i'j'}}\,\Delta_{ij}\,\Delta_{i'j'}\,\overline{\Delta}_{iji'j'}
<\frac{1}{(2^N+1)\,4^N}\sum_{i\neq j(\neq 0_v)}3^{r_{ij}}=\frac{{\Vert P\Vert}^2}{2^N+1}<\bigg(\frac{3}{4}\bigg)^{\!N}.
\end{align}
Equation (\ref{smaller than 3/4}) shows that, for any $P$, $\var(R[P])$ for a random state is bounded by ${\Vert P\Vert}_2^2<1$ with probability exponentially close to 1 as a function of $N$.

%%%%%%%%%%%%%%%%%%%%%%%%%%%%%%%%%%%%%%%%%%%%%%
%                                                           APPENDIX 3 - PAULI TOMOGRAPHY															 %
%%%%%%%%%%%%%%%%%%%%%%%%%%%%%%%%%%%%%%%%%%%%%%

\subsection{Pauli tomography} \label{pauli}
\noindent

Equation (\ref{tomo N-qubit}) can be derived taking the trace of (\ref{general O}) after having multiplied both members by the density operator and calculating the trace on the eigenvectors of $\sigma_{i_1}\cdots\sigma_{i_N}$:
\begin{align} \label{pauli tomo expanded}
{\langle O\rangle}&={\mathrm {Tr}}\big[O\,\rho\big]=\sum_i \,a_i\,{\mathrm {Tr}}\big[O\,\sigma_{i_1}\cdots\sigma_{i_N}\big]=\sum_{i_1}\cdots\sum_{i_N}\sum_{m_{i_1}}\cdots\sum_{m_{i_N}}\,p(m_{i_1},\ldots,m_{i_N})\,a_i\,m_{i_1}\cdots m_{i_N}\\\nonumber&=\bigg(\frac{1}{4^N}\prod_{k=1}^N \,\sum_{i_k=0}^{3}\,\sum_{m_{i_k}}\bigg)\,p(m_{i_1},\ldots,m_{i_N})\,4^N\,a_i\,m_{i_1}\cdots m_{i_N}.\nonumber						
\end{align}
The same results could have been obtained from (\ref{expectation values}) choosing the dihedral subgroup of $SU(2)$ as $\cal G$~\cite{Paini2000Quantum},~\cite{DAriano2003Spin}.

\par
\vskip 1em

As explained in Section \ref{SU(2) special}, the Monte Carlo method applied to (\ref{pauli tomo expanded}) is ineffective due to the likely high number of projections equal to zero of the components of $O$ of (\ref{general O}) on the sampled vectors $\{\sigma_{j_1}\cdots\sigma_{j_N}\}$. However, for a given $O$, we can simplify (\ref{pauli tomo expanded}) and only sum on values of $i$ for which the coefficients $a_i$ are different from zero. Let $L_O$ be the set of such values of $i$ and $\vert L_O\vert$ its cardinality. Equation (\ref{pauli tomo expanded}) becomes
\begin{align} \label{only on L_O}
{\langle O\rangle}&\!=\!\bigg(\frac{1}{4^N}\prod_{k=1}^N \sum_{i_k=0}^{3}\sum_{m_{i_k}}\bigg)p(m_{i_1},\ldots,m_{i_N})\,4^N\,a_i\,m_{i_1}\cdots m_{i_N}
\!\equiv\!\bigg(\frac{1}{4^N}\sum_i\sum_{m_i}\bigg)p(m_{i_1},\ldots,m_{i_N})\,4^N\,a_i\,m_{i_1}\cdots m_{i_N}
\\\nonumber&\!=\!\bigg(\frac{1}{\vert L_O\vert}\sum_{i\in L_O}\sum_{m_i}\bigg)p(m_{i_1},\ldots,m_{i_N})\,\vert L_O\vert\,a_i\,m_{i_1}\cdots m_{i_N}\!\equiv\!\bigg(\frac{1}{\vert L_O\vert}\sum_{i\in L_O}\sum_{m_i}\bigg)p(m_{i_1},\ldots,m_{i_N})\,R'[O](m_{i_1},\ldots,m_{i_N})\nonumber						
\end{align}
and no longer has the problem of the null projections. The term $p(m_{i_1},\ldots,m_{i_N})$ represents the probability of obtaining $(m_{i_1},\ldots,m_{i_N})$ measuring $\sigma_{i_1},\cdots,\sigma_{i_N}$ for a given $i$ and $1/{\vert L_O\vert}$ the normalisation of the sum of the probabilities, assuming uniform sampling of $i$ from $L_O$. If we generalise to sampling $i$ from the probability $\tilde {p}(i)$ defined on $L_O$, Eq. (\ref{only on L_O}) becomes
\begin{align}
{\langle O\rangle}&=\sum_{i\in L_O}\,\sum_{m_i}\,\tilde {p}(i)\,p(m_{i_1},\ldots,m_{i_N}|i)\,\tilde {R}[O](m_{i_1},\ldots,m_{i_N}),						\label{with tilde}
\end{align}
with $\tilde {R}[O](m_{i_1},\ldots,m_{i_N})={\tilde {p}}^{-1}(i)\,a_i\,m_{i_1}\cdots m_{i_N}$. Equation (\ref{variance}) tells us that the effectiveness of a tomographic procedure, given the same number of preparations, only depends on the variance of the estimator. We can therefore try to choose $\tilde {p}$ to minimise the variance, or, at least, to minimise the expectation value of the square of the estimator, which is an upper bound for the variance. If we assume ${\tilde {p}}(i)$ has the form
\begin{align}
{\tilde {p}}_x(i)=\frac{|a_i|^x}{\sum_{j\in L_O}	\,|a_j|^x},		
\end{align}
with $x$ positive real number, the expectation value of the square of the estimator is
\begin{align}
\langle{\tilde {R}}^2[O]\rangle(x)&=\sum_{i\in L_O}\,\sum_{m_i}\,{\tilde {p}}_x(i)\,p(m_{i_1},\ldots,m_{i_N}|i)\,{\tilde {R}}^2[O](m_{i_1},\ldots,m_{i_N})=\sum_{i\in L_O}\,\frac{|a_i|^2}{{\tilde {p}}_x(i)}\,\sum_{m_i}\,p(m_{i_1},\ldots,m_{i_N}|i)\\
&=\sum_{i\in L_O}\,|a_i|^{2-x}\,\sum_{j\in L_O}\,|a_j|^{x},  \nonumber
\end{align}
with minimum in $x=1$. We choose $\tilde {p}(i)\!=\!{{\tilde {p}}_1(i)}$ and obtain an estimator with variance bounded 
by ${(\sum_{i}\,|a_i|)}^2$, as in the quantum expectation estimation procedure~\cite{MRBA,WHT,RBM}.

%%%%%%%%%%%%%%%%%%%%%%%%%%%%%%%%%%%%%%%%%%
%                                                           APPENDIX 4                                                                %
%%%%%%%%%%%%%%%%%%%%%%%%%%%%%%%%%%%%%%%%%%%%%%

\subsection{Seminorm changes under unitary transformations} \label{norm changes}
\noindent

We consider the linear operator $\cal U$ from ${\cal B}(\cal H)$ to ${\cal B}(\cal H)$, defined as ${\cal U}(O)=U^{\dagger}OU$, with $O,U\in{\cal B}(\cal H)$ and $U$ unitary in $\cal H$. With the linear product $(O_1,O_2)\equiv\tr[O_1^{\dagger}\,O_2]$, it is easy to check that $\cal U$ is unitary in ${\cal B}(\cal H)$. If 
\begin{align}
O=\sum_i \,a_i\,\sigma_{i_1}\cdots\sigma_{i_N}  \label{O expansion again}
\end{align}
as in (\ref{general O}) and 
\begin{align}
{\cal U}(O)=\sum_i \,b_i\,\sigma_{i_1}\cdots\sigma_{i_N},  
\end{align}
then the unitarity of ${\cal U}$ implies that
\begin{align}
\sum_i \,a_i^2=\sum_i \,b_i^2.  \label{length maintained}
\end{align}
The linear space ${\cal B}(\cal H)$ can be expressed as the direct sum of subspaces ${{\cal B}(\cal H)}={\oplus_{k=0}^N \,{\cal B}_k}$, with ${\cal B}_k$ representing the subspace spanned by the Pauli monomials $\{\sigma_{i_1}\cdots\sigma_{i_N}\}$ of the same degree in Pauli's operators, or, equivalently, with the same number of indices $i_l\neq0$ in $i=({i_1},\ldots,{i_N})$. If we write $O$ of (\ref{O expansion again}) as the sum of its components on these subspaces
\begin{align}
O=\sum_{k=0}^N\bigg(\sum_{i'}\,a_{ki'}\,\sigma_{(ki')_1}\cdots\sigma_{(ki')_N}\bigg),  \label{O again}
\end{align}
with $i'$ enumerating the Pauli monomials in each subspace, the seminorm (\ref{original norm}) of $O$ becomes
\begin{align}
{\Vert O\Vert}_2^2=\sum_{k=1}^N\,3^{k}\,\bigg(\sum_{i'}\,a_{ki'}^2\bigg).							\label{norm N qubits new}
\end{align}

Identity operators remain unchanged by the application of any ${\cal U}$. If $U$ is the product of single--qubit operators as in (\ref{U as single unitaries}), since $U_1^{\dagger} \sigma_i U_1$ for every $i\neq0$ is a linear combination of Pauli operators, every subspace ${\cal B}_k$ is invariant by ${\cal U}$ and
\begin{align}
\sum_{i'} \,a_{ki'}^2=\sum_{i'} \,b_{ki'}^2\;\;\forall k,		
\end{align}
which implies ${\Vert O\Vert}_2^2={\Vert\,{\cal U}(O)\Vert}_2^2$.

If $U$ is the product of 2--qubit operators as in (\ref{U as double unitaries}), the case is slightly more complicated. We start from a single $U_2$ of (\ref{U as double unitaries}), which we assume to act on qubits 1 and 2 for simplicity. Clearly ${\cal B}_0$ is still invariant by $\cal U$, but the other subspaces are not. In fact, if we consider the system of qubits 1 and 2 only, we can choose for example the 2--qubit operator $\mathbb{1}\otimes\sigma_z$, which transforms into $\sigma_x\otimes\sigma_y$ under $\cal U$, with $U=e^{i(\pi/4)\sigma_x\otimes\sigma_x}$. For the system of the 2 qubits, given any operator can be expressed as the sum of its 3 components in ${\cal B}_0, {\cal B}_1$ and ${\cal B}_2$ and given the invariance of ${\cal B}_0$ for $\cal U$, we can only conclude that ${\cal B}_1\oplus{\cal B}_2$ is invariant for $\cal U$, but not ${\cal B}_1$ and ${\cal B}_2$ separately. Because of (\ref{norm N qubits new}), the square of the seminorm in (\ref{norm N qubits new}) of an operator in ${\cal B}_1\oplus{\cal B}_2$ will change by a factor between $1/3$ and 3 after the application of ${\cal U}$, where the limit cases are represented by operators of ${\cal B}_2$ mapped to operators of ${\cal B}_1$ by ${\cal U}$ and vice versa. Coming back to the system of $N$ qubits, the application of $U_2$ will, as a result, increase the square of the seminorm (\ref{norm N qubits new}) by up to a factor 3. If $N/2$ operators $U_2$ are applied as in (\ref{U as double unitaries}) and if in the expansion of $O$ there is a term containing a Pauli operator for every pair $\{1,2\},\{3,4\},\ldots,\{N-1,N\}$, then each $U_2$ will contribute to increasing the square of the seminorm (\ref{norm N qubits new}) by a factor at most given by 3. More in general, if $P$ is the minimum between the highest power of the Pauli monomials of $O$ and $N/2$, then the seminorm (\ref{norm N qubits new}) will increase at most by a factor $3^{P/2}$ with a transformation $U$ as in (\ref{U as double unitaries}) (although, it could also decrease by up to a factor of $3^{-P/2}$).

\end{widetext}

\end{document}